\documentstyle[prc,preprint,tighten,aps]{revtex}
\begin {document}
\draft
\title{
Parity-violating$\mbox{\boldmath{$\alpha$}}$-decay
of the 3.56-MeV ${\bf J^\pi,T=0^+,1}$ \\
state of \bbox{^6}Li}
\author{Attila Cs\'ot\'o$^1$ and Karlheinz Langanke$^2$}
\address{$^1$National Superconducting Cyclotron Laboratory,
Michigan State University, East Lansing, Michigan 48824 \\
$^2$W.~K. Kellogg Radiation Laboratory, California
Institute of Technology, Pasadena, California 91125}
\date{November 15, 1995}

\maketitle

\begin{abstract}
\noindent
We study the parity-violating $\alpha +d$ decay of the
lowest $0^+$ state of $^6$Li in a microscopic
three-cluster model. The initial bound and the final
scattering states are described consistently within the
same model. The parity-violating decay width is calculated
in perturbation theory using the parity-nonconserving
nucleon-nucleon interaction of Desplanques, Donoghue, and
Holstein (DDH). We find that the decay width is sensitive
to dynamical degrees of freedom which are beyond the
$\alpha+p+n$ model, for example, $\alpha$ excitation and
breakup. In our full model, which contains breathing
excitations of the $\alpha$ particle and $^3$H+$^3$He
rearrangement, the parity-nonconserving decay width is
$\Gamma_{\alpha d}^{PNC}=3.97\cdot 10^{-9}$ eV, using
the DDH coupling constants.
\end{abstract}

\narrowtext

\section{Introduction}

Parity violation has played an important role in the
understanding of the nature of the weak interaction.
Today the weak interaction is well understood in the
leptonic, semileptonic, and strange nonleptonic sectors.
However, our picture of the nonstrange nonleptonic
weak interaction, which appears in nuclear processes,
e.g.\, in $n+p$ scattering, is far from being complete
\cite{Adelberger,Holstein}. The presence of the strong
part of the nucleon-nucleon force makes the effect of
the weak nucleon-nucleon interaction hardly observable.
The size of the parity violating effect is, for example,
typically in the order of $10^{-7}$ relative to the
effect of the strong $N$-$N$ force.

Despite the smallness of its effect, parity violation
in a nuclear process was experimentally observed in
1967 by Lobashov {\it et al.} \cite{Lobashov}. Since
then, nuclear parity violation has been studied in $p+p$
scattering and $p+n$ capture, in few-nucleon systems (e.g.\
$p+d$ and $p+\alpha$ scattering), in parity-mixed
doublets of light nuclei, and in polarized neutron
scattering on heavy nuclei. For two excellent review
articles on the nuclear parity violation experiments,
see Refs.\ \cite{Adelberger,Holstein}.

Complex nuclear structure and dynamics often leads to
ambiguities in the theoretical description of nuclear
parity violation involving light nuclei. However,
impressive theoretical progress has been achieved recently
in the description of the six-nucleon systems
\cite{GFMC,Zheng,Tang,He6,Kok,Li6ref,Kukulin}. While
realistic GFMC results for light nuclei, including for
$^6$Li, are very encouraging \cite{GFMC}, these studies
are basically restricted to bound states. However,
effective many-body theories like the microscopic
multicluster model yield a good and consistent description
of the nuclear structure and dynamics (bound and
scattering states) in the six-nucleon systems simultaneously
\cite{Tang,He6}. This method appears thus to be well
suited for the study of the parity-violating $\alpha +d$
decay of the $J^\pi,T=0^+,1$ state of $^6$Li at 3.56 MeV
excitation energy. As the spin-parity of the deuteron is
$1^+$, the decay of the $0^+$ $^6$Li state into the
$\alpha +d$ channel is only possible if  the continuum
final state is $J^\pi=0^-$; thus this process violates
parity. As the final state is $T=0$, this process is
sensitive to the isovector part of the parity-violating
$N$-$N$ potential.
The best experimental upper limit for the parity-violating
$\alpha +d$ partial decay width is $\Gamma \leq 6.5\cdot
10^{-7}$ eV \cite{Robertson}. The two most comprehensive
theoretical descriptions of this process are based on the
shell model \cite{Brown} and the harmonic oscillator
cluster model \cite{Burov}, respectively. If one assumes
the same parameters for the parity-violating $N$-$N$
potential, these two studies find parity-violating decay
widths which  differ by orders of magnitude.
Both of these models contain questionable approximations.
In \cite{Burov} the initial $0^+$ wave function is rather
schematically described by one $\alpha (pn)$ configuration
with total spin and angular momentum zero.
Note that the initial state has a spatially extended
halo-like neutron-proton tail \cite{Varga}.
The ability of the shell model \cite{Brown} to reproduce
such a loosely bound state with genuine three-body
($\alpha+p+n$) nature is questionable. A model which cannot
reproduce this tail tends to compress the wave function
inside, thus supposedly increasing the decay width.
Moreover, both studies \cite{Brown,Burov} use a potential
model for the $\alpha +d$ scattering, which is inconsistent
with the description of the bound state.

In the present paper we study the problem in a six-body
three-cluster model, which is practically complete in the
$\alpha +p+n$ space. We also investigate the effects of
$\alpha$ distortion and t+$^3$He rearrangement on the decay
width. Our model correctly describes the spatially extended
nature of the $0^+$ bound state, and uses an $\alpha +d$
scattering state which is consistent with the bound state.
We try to use a model which is as parameter-free as
possible by requiring the good reproduction of the
properties of the subsystems ($N+N$ and $\alpha +N$
scattering states, channel thresholds, binding energies,
etc.).

\section{Model}

We use the parity-nonconserving nucleon-nucleon potential
of Desplanques, Donoghue, and Holstein \cite{DDH}, which
was derived from a valence quark model. The isovector
part of this potential reads
\widetext
\begin{eqnarray}
V^{PNC}(r)&=&{{f_\pi g_{\pi NN}i}\over{\sqrt{2}}}
\left ( {{{\mbox{\boldmath $\tau$}}_1\times
{\mbox{\boldmath $\tau$}}_2}\over{2}}\right )_z
({\mbox{\boldmath $\sigma$}}_1+{\mbox{\boldmath
$\sigma$}}_2)\cdot \left [ {{{\bf p}_1-{\bf p}_2}\over{2M}},
f_\pi(r) \right ] \cr
&-&g_\rho h_\rho^1\left ({{{\mbox{\boldmath $\tau$}}_1
+{\mbox{\boldmath $\tau$}}_2}\over{2}}\right )_z
\Bigg (({\mbox{\boldmath $\sigma$}}_1-{\mbox{\boldmath
$\sigma$}}_2)\cdot \left \{{{{\bf p}_1-{\bf p}_2}\over{2M}},
f_\rho(r) \right \} \cr
&& \hskip 2cm
+i(1+\chi_V)({\mbox{\boldmath
$\sigma$}}_1\times {\mbox{\boldmath $\sigma$}}_2)\cdot
\left [{{{\bf p}_1-{\bf p}_2}\over{2M}},f_\rho(r) \right ]
\Bigg ) \cr
&-&g_\omega h_\omega^1\left ({{{\mbox{\boldmath $\tau$}}_1
+{\mbox{\boldmath $\tau$}}_2}\over{2}}\right )_z
\Bigg (({\mbox{\boldmath $\sigma$}}_1-{\mbox{\boldmath
$\sigma$}}_2)\cdot \left \{{{{\bf p}_1-{\bf p}_2}\over{2M}},
f_\omega(r) \right \} \cr
&& \hskip 2cm
+i(1+\chi_S)({\mbox{\boldmath
$\sigma$}}_1\times {\mbox{\boldmath $\sigma$}}_2)\cdot
\left [{{{\bf p}_1-{\bf p}_2}\over{2M}},f_\omega(r) \right ]
\Bigg ) \cr
&+&g_\rho h_\rho^1
\left ( {{{\mbox{\boldmath $\tau$}}_1-
{\mbox{\boldmath $\tau$}}_2}\over{2}}\right )_z
({\mbox{\boldmath $\sigma$}}_1+{\mbox{\boldmath
$\sigma$}}_2)\cdot \left \{ {{{\bf p}_1-{\bf p}_2}\over{2M}},
f_\rho(r) \right \} \cr
&-&g_\omega h_\omega^1
\left ( {{{\mbox{\boldmath $\tau$}}_1-
{\mbox{\boldmath $\tau$}}_2}\over{2}}\right )_z
({\mbox{\boldmath $\sigma$}}_1+{\mbox{\boldmath
$\sigma$}}_2)\cdot \left \{ {{{\bf p}_1-{\bf p}_2}\over{2M}},
f_\omega(r) \right \},
\label{PNC}
\end{eqnarray}
\narrowtext
\noindent
where $r=\vert {\bf r}_1-{\bf r}_2 \vert$,
$f_\alpha(r)=\exp (-m_\alpha r)/4\pi r$ with $m_\alpha$
being the $\pi^\pm$, $\rho$, and $\omega$ meson masses,
${\bf p}_i$ are the nucleon momenta, $M$ is the nucleon
mass, and $\mbox{\boldmath $\sigma$}$ and $\mbox{\boldmath
$\tau$}$ are the spin and isospin Pauli matrices,
respectively. The scalar and vector magnetic moments are
$\chi_S=-0.12$, and $\chi_V=3.70$, and [...] and \{...\}
denote commutators and anticommutators, respectively. We
use the redefined coupling constants of \cite{Adelberger}
\begin{equation}
F_\pi=f_\pi g_{\pi NN}/\sqrt{32}, \; F_1=-g_\rho
h_\rho^1/2, \; G_1=-g_\omega h_\omega^1/2.
\end{equation}
The best values and reasonable ranges of the coupling
constants, based on the DDH theory, can be found in
\cite{Adelberger,DDH}.

The parity-violating $\alpha +d$ decay width is
calculated perturbatively
\begin{equation}
\Gamma_{\alpha d}^{PNC}=\hbar W_{fi}=2\pi \Big \vert
\langle \Psi^{\alpha d} \vert {\widehat V}^{PNC}
\vert \Psi^{^6{\rm Li}} \rangle \Big \vert ^2
\varrho (E_f),
\label{width}
\end{equation}
where $\varrho (E_F)$ is the density of the continuum
states at the final state energy $E_f=(3.563-1.475)=2.088$
MeV.

Our initial $^6$Li state reads
\widetext
\begin{eqnarray}
\Psi^{^6{\rm Li}}=\Psi^{\alpha pn}+\Psi^{th}
&=&\sum_{(ij)k,S,l_1,l_2,L}
{\cal A}\left \{\left [ \left [\Phi^i(\Phi ^j\Phi^k)
\right ]_S
\chi ^{i(jk)}_{[l_1l_2]L}(\mbox{\boldmath
$\rho $}_{jk},\mbox{\boldmath $\rho $}_{i(jk)})
\right ]_{JM} \right \} \cr
&+&\sum_{S,L} {\cal A}\left \{\left [ [\Phi ^t\Phi ^h
]\raise-0.66ex\hbox{\scriptsize $S$}
\chi ^{th}_L(\mbox{\boldmath $\rho $}_{th})
\right ]_{JM} \right \},
\label{Li6}
\end{eqnarray}
\narrowtext
\noindent
where the indices $i,j$, and $k$ denote any one of the
labels $\alpha$, $p$, and $n$. In (\ref{Li6}) ${\cal A}$
is the intercluster antisymmetrizer, the $\Phi$ cluster
internal states are translationally invariant harmonic
oscillator shell model states, the \mbox{\boldmath
$\rho $} vectors are the different intercluster Jacobi
coordinates, and [...] denotes angular momentum coupling.
The sum over $S,l_1,l_2$, and $L$ includes all
angular momentum configurations of any significance.
The last term in (\ref{Li6}) is the $t+h$=$^3$H+$^3$He
rearrangement channel. The monopole breathing distortions
of the $\alpha$ particle is considered by
\begin{equation}
\Psi^{\alpha pn}=\Psi^{\alpha_1pn}+\Psi^{\alpha_2pn}+...,
\label{Li61}
\end{equation}
where the antisymmetrized ground state ($i=1$) and
monopole excited states ($i>1$) of the $\alpha$ particle
are represented by the wave functions
\begin{equation}
\Phi ^{\alpha_i}=
\sum_{j=1}^{N_\alpha}A_{ij}\phi ^\alpha _{\beta _j},\ \
i=1,2,...,N_\alpha.
\label{alpha}
\end{equation}
Here $\phi ^\alpha _{\beta _j}$ is a translationally
invariant shell--model wave function of the $\alpha$
particle with size parameter $\beta _j$ and the $A_{ij}$
parameters are to be determined by minimizing the energy
of the $\alpha$ particle \cite{Tang}. We choose the same
parameters for the wave function as in Ref. \cite{He6}.
We note, that the same model excellently
reproduced the neutron halo structure of $^6$He
\cite{He6}. Thus the current approach is
adequate to describe spatially extended systems, like the
$0^+$ state of $^6$Li, which is the analog of the
$^6$He ground state.

The wave function of the final continuum state reads
(with $L=1$, $S=1$, and $J^\pi=0^-$),
\begin{eqnarray}
\Psi^{\alpha d}
&=&{\cal A}\left \{\left [ [\Phi
^{\alpha_1} \Phi ^{d_1}]\raise-0.66ex\hbox{\scriptsize $S$}
\ g^{\alpha_1 d_1}_{L}(E,\mbox{\boldmath $\rho $}_{
\alpha d})\right ]_{JM}\right \}\nonumber \\
&+&
\sum_{i=2}^{N_\alpha}
\sum_{j=2}^{N_d}{\cal A}\left \{\left [ [\Phi
^{\alpha_i}\Phi ^{d_j}]\raise-0.66ex\hbox{\scriptsize $S$}
\ \chi^{\alpha_i d_j}_L(E,\mbox{\boldmath $\rho $}_{\alpha
d})\right ]_{JM}
\right \}\nonumber \\
&+&{\cal A}\left \{\left [ [\Phi ^t\Phi ^h
]\raise-0.66ex\hbox{\scriptsize $S$}
\ \chi^{th}_L(E,\mbox{\boldmath $\rho $}_{th})
\right ]_{JM} \right \}.
\label{ad}
\end{eqnarray}
Here $E$ is the $\alpha +d$ relative motion energy in the
center-of-mass frame. To account for specific distortion
effects in the deuteron, we expand its wave function by
5 basis states ($N_d=5$).
The ground state $\Phi ^{d_1}$ reproduces the deuteron
binding energy ($-2.20$ MeV) and (point nucleon) rms-radius
(1.95 fm) very well. In (\ref{width})  the $\alpha +d$
final state is taken to be the time-reversed of a state
with an incoming plane wave, $\exp (i{\bf k r})$, and
scattered spherical waves. The plane wave is then projected
to $L=1$. The normalization of $g$ in (\ref{ad}) is chosen
consistently with the form of the plane wave. Thus for
$\rho_{\alpha d}\rightarrow\infty$ one has
\begin{eqnarray}
g^{\alpha_1 d_1}_{1}(E,\mbox{\boldmath $\rho$}_{\alpha d})
&\rightarrow & Y_{1m} (\hat{\mbox{\boldmath
$\rho $}}_{\alpha d}) (k\rho_{\alpha d})^{-1} \nonumber \\
&\times& \Big (F_1(k\rho_{\alpha d})\cos \delta
+G_1(k\rho_{\alpha d})\sin \delta \Big ) ,
\end{eqnarray}
where $k$ is the
wave number,
$F_1$ and $G_1$ are Coulomb functions,
and $\delta$ is the $^3$$P_0$ $\alpha +d$ phase shift
at energy $E$.

Our model wave functions contain a large and physically
most important part of the six-body Hilbert space.
Although, our model is currently probably the closest
approximation to a consistent and dynamically correct full
six-body description of the parity-violating decay
process, it still has limitations. To estimate these
limitations we perform a series of calculations
in increasingly sophisticated model spaces by subsequently
adding $t+h$ rearrangement and $\alpha$ distortions  to
our model.

Putting (\ref{Li6})-(\ref{Li61}) into the six-nucleon
Schr\"odinger equation which contains a parity-conserving
two-nucleon strong and Coulomb interaction, we arrive at
an equation for the intercluster relative motion functions
$\chi$. These functions are expanded in terms of products
of tempered Gaussian functions $\exp (-\gamma_i \rho^2)$
\cite{Kamimura} with different ranges $\gamma_i$ for each
type of relative coordinate. The expansion coefficients
are determined from a variational principle. The
scattering states are calculated from a Kohn-Hulth\'en
variational method for the $S$-matrix, which uses square
integrable basis functions matched with the correct
scattering asymptotics \cite{Kamimura}. Then, using the
resulting six-nucleon wave functions, the decay width
(\ref{width}) is evaluated. All necessary matrix elements
are calculated analytically by the aid of a symbolic
computer language \cite{reduce}.

\section{Results}

We consider four different model spaces with
increasing level of sophistication:\ (i) no $\alpha$
breathing modes ($N_\alpha=1$) and no $t+h$ rearrangement
channel: $\{\alpha_1+p+n\}$; (ii) $N_\alpha=1$
and the $t+h$ channel is included: $\{\alpha_1+p+n;t+h
\}$; (iii) $N_\alpha=3$ and no $t+h$ channel:
$\{\alpha_3+p+n \}$; (iv) $N_\alpha=3$ and the $t+h$
channel is included: $\{\alpha_3+p+n;t+h \}$.
We use the Minnesota effective interaction \cite{MN}
as the parity-conserving strong nucleon-nucleon force.
It has been shown \cite{He6}, that this interaction
provides excellent $N+N$ and $\alpha +N$ phase shifts in
the $\{ \alpha_3+N,T+d \}$ and $\{\alpha_1+N \}$ models
by using $u=0.92$ and $u=0.98$ exchange mixture
parameters of the central interaction and slightly
different spin-orbit forces, respectively.
Here we refit the $u$ parameter in order to reproduce the
experimental binding energy of the $0^+$ state (0.137 MeV
relative to the $\alpha+p+n$ threshold) in all four
models. These adjustments are rather small. For example,
in our full model we have to change $u$ by only 0.7\%.
The weights of the $(L,S)=(0,0)$ and $(1,1)$ components of
the wave functions are between 86.5--89.5\% and
13.5--10.5\%, respectively in the four different models in
accordance with the results of several different model
calculations (e.g., \cite{Li6ref,Kukulin}).

In Fig.\ \ref{fig1} we show the $0^-$ phase shifts of our
four models, together with the experimental data
\cite{Jenny} and the phase shift generated by the
McIntyre-Haeberli optical potential \cite{MH}. Note that
the two lowest-energy experimental data points are not
tabulated in \cite{Jenny}, we read them off the figure of
Ref.\ \cite{Jenny}. All our theoretical models slightly
underestimate the absolute value of the phase shift at
the $E=2.09$ MeV, which is the final energy of the decay
process. The full model (iv) is closest to experiment.
Below we investigate the sensitivity of the decay width
on the phase shift at 2.09 MeV.

The low-lying $\alpha +d$ spectrum of $^6$Li is reproduced
well by all four models. For example, in the
$\{\alpha_1+p+n,t+h\}$ model, the $J^\pi,T$=$1^+,0$
(ground state), $3^+,0$, and $2^+,0$ $^6$Li states are
at --1.421 MeV, 0.637 MeV ($\Gamma=0.012$ MeV), and 4.254
MeV ($\Gamma=2.78$ MeV), respectively, while the
experimental values are --1.475 MeV, 0.71 MeV
($\Gamma=0.024$ MeV), and 2.83 MeV ($\Gamma=1.7$ MeV).
All energies are relative to the $\alpha +d$ threshold.
As in all models the deuteron has the correct
binding energy, the position of the $\alpha +d$
threshold relative to the $\alpha +p+n$ is always
reproduced. Moreover, the $^5$He+$p$ and $^5$Li+$n$
thresholds are at the correct position, while our $t+h$
threshold is 4-5 MeV higher than experimentally.

In Table \ref{tab1} we show the parity-violating $\alpha
+d$ partial decay widths given by our four different model
spaces, taking into account only the pion term of
(\ref{PNC}) and using the DDH best value for $F_\pi$. The
spread among the results is almost an order of magnitude.
To understand the origin of these big differences
we performed some test calculations. We found that the
width depends moderately on the $\alpha +d$ phase shift
at 2.09 MeV. If we used a scattering wave function which
reproduced the phase shift of the McIntyre-Haeberli
potential at 2.09 MeV, then
$\Gamma_{\alpha d}^{PNC}$ was increased by roughly 50\%.
We checked the sensitivity of $\Gamma$ on the binding
energy of the $0^+$ state by using the original $u=0.98$
exchange mixture parameter in our simplest model. The
binding energy decreases by 350 keV (resulting in a
positive energy pseudo-bound state) which leads to a 15\%
decrease in $\Gamma$. Obviously, a larger probability of
finding nucleons farther from each other  decreases
somewhat the magnitude of the internal wave function and
reduces $\Gamma$.

We checked the role of the orthogonal $(L,S)=(0,0)$ and
(1,1) components in the bound state wave function
of our four models by performing calculations for the
two components separately. In each case the exchange
mixture parameter $u$ has been refitted to reproduce
the correct $0^+$ binding energy. For the
$\{ \alpha_1+p+n\}$ model space, the individual
$(0,0)$ and $(1,1)$ components yield widths of $2.13\cdot
10^{-10}$ eV and $2.23\cdot 10^{-8}$ eV, respectively.
This is to be compared with the result $\Gamma=1.06\cdot
10^{-9}$ eV, obtained if both components are considered.
Although we observe a strong sensitivity of $\Gamma$ on
the weight of the $(L,S)=(1,1)$ component, this cannot
explain the big differences in Table \ref{tab1} alone, as
the weight of this component is roughly the same in all
model spaces.

In Table \ref{tab2} we list the partial contributions of
the $(L,S)=(0,0)$ and $(1,1)$ components of the bound
state to the parity-violating matrix element for
our four model spaces. One can see that the contribution
of the $(0,0)$ component changes its sign when going from
the model with only one $\alpha$-particle basis state
to the one which contains $\alpha$ breathing modes. This
is obviously the consequence of some kind of cancellations
taking place in the matrix element. These strong
cancellations appear to be a warning that it is dangerous
to generate bound and scattering states inconsistently
from different models, as has been done in \cite{Brown}
and \cite{Burov}. A similar sensitivity is found in the
modelling of the beta-delayed deuteron emission process
in $^6$He \cite{beta}.

As $\Gamma$ only moderately depends on the scattering
state, the main origin of these cancellations must
be in the bound state wave function. As the
matrix element of the parity-violating interaction cannot
be rewritten in a form which contains only relative
motions, we cannot directly determine which property of
the bound state wave function causes the sign change in
the $(0,0)$ component. By investigating the contribution
of the individual $\alpha$-particle basis states, we find
that the contributions of the excited $\alpha$
pseudo-states are substantial, but they do not cause the
sign change. As in Ref.\ \cite{beta}, their effect might be
to shift the nodal positions in the bound state wave
function.

{}From Table II we observe that our model has a
non-anticipated sensitivity to the model space, showing
that the parity-violating decay is determined by components
in the model space which are not well constrained by the
other low-energy properties of the 6-nucleon systems which
our models describe quite well. Nevertheless we will
complete our study by calculating the decay width $\Gamma$
within our most elaborate model space (iv) and using the
full parity-violating interaction as defined in Eqs.\
(1)-(2). Then
\begin{equation}
\Gamma_{\alpha d}^{PNC}=(55.2\cdot F_\pi+9.26\cdot
F_1+6.59\cdot G_1)^2\;\; {\rm eV}.
\end{equation}
In Table \ref{tab3} we give $\Gamma$ for the coupling
constants of \cite{DDH}, \cite{DZ}, and \cite{FCDH},
respectively. One observes about an order of magnitude
spread among the decay widths calculated for these
different sets of coupling constants.
For comparison, Table \ref{tab3} also gives $\Gamma$
using only the pion contribution in the parity-violating
potential. As the $\rho$ and $\omega$ contributions
are small, the decay width shows about an order
of magnitude variation among the various model spaces.

Finally, we compare our results with the predictions of
\cite{Brown} and \cite{Burov}. In these models the pionic
term alone gives $1.3\cdot 10^{-8}$ eV and $1.1\cdot
10^{-11}$ eV, respectively, using the DDH best
value for the coupling constant. In \cite{Burov} only
the $(L,S)=(0,0)$ component is included in the bound
state wave function which might partly explain the
smallness of their reported decay width. In \cite{Brown}
the $(L,S)$ weights are similar to ours, being 89.4\% and
10.4\% for $(0,0)$ and $(1,1)$, respectively. Compared to
\cite{Brown}, we do not take into account the $D$ state of
the deuteron (our parity-conserving $N$-$N$ interaction is
designed for a pure $S$-state deuteron). According to
\cite{Brown}, inclusion of the deuteron D-state
increases $\Gamma$ by 50\%. Thus our decay width appears
to be about a factor of two smaller than the estimate
given in Ref. \cite{Brown} which is partly due to the use
of a too attractive scattering potential in \cite{Brown}.
In contrast to the present results, Ref.\ \cite{Burov}
finds a substantial contribution of the $\omega$ term in
the parity-violating interaction to the decay width.

\section{Conclusion}

In summary, we have studied the parity-violating $\alpha+d$
decay of the lowest $0^+$ state of $^6$Li within a
microscopic multicluster model. We have performed a series
of studies based on a three-cluster $\alpha +p+n$ model
space (including all possible arrangement of the clusters
and within each arrangement all relevant angular momenta)
and its extension to additionally include monopole
breathing modes of the $\alpha$ particle and the
$^3$H+$^3$He rearrangement channel. For the
parity-conserving interaction we used the Minnesota force
which reproduces all relevant subsystem properties
reasonably well. We have calculated the decay width in
perturbation theory, using a consistent description of
the $0^+$ bound state and the final $0^-$ scattering state.

We have found that the parity-nonconserving $\alpha +d$
decay width $\Gamma_{\alpha d}^{PNC}$ is moderately
sensitive to the correct reproduction of the experimental
$\alpha +d$ phase shift and the $0^+$ binding energy.
On the other hand, we have demonstrated that
$\Gamma_{\alpha d}^{PNC}$ is very sensitive to the presence
of the 10--15\% $(L,S)=(1,1)$ component in the bound state
wave function. As the inclusion of $\alpha$ breathing
modes and the $^3$H+$^3$He rearrangement channel changes
$\Gamma$ considerably, we find that dynamical degrees of
freedom beyond the $\alpha +p+n$ three-cluster space are
quite important. The decay width in our most complete
model, using the DDH best values for the weak coupling
constants, is $\Gamma_{\alpha d}^{PNC} = 3.97\cdot
10^{-9}$ eV compared to the experimental upper limit
$\Gamma_{\alpha d}^{PNC}\leq 6.5\cdot 10^{-7}$ eV.

Our most important result, however, is that the
state-of-the-art microscopic multicluster model does not
constrain the parity-violating decay width of the $T=1$
state at $E=3.56$ MeV in $^6$Li sufficiently well; despite
the fact that this model is very successful in
simultaneously describing other low-energy properties of
the six-nucleon system. We note that a similar
sensitivity has already been observed in cluster model
studies of the beta-delayed deuteron emission process in
$^6$He \cite{beta}. Improved dynamical calculations in
larger six-body model spaces would be welcome to clarify
the role of other degrees of freedom beyond our model, for
example, by including the D-state components in the
deuteron and $\alpha$-particle. Such calculations are,
however, very challenging, as they have to consistently
reproduce the spatially extended halo-like nature of the
$^6$Li $0^+$ state and the $\alpha +d $ scattering state.

\acknowledgments

The work of A.\ C.\ was supported by Wolfgang Bauer's
Presidential Faculty Fellowship (PHY92-53505). This work
was also supported by NSF Grant No.\ PHY94-03666 (MSU) and
PHY94-15574 and PHY94-12818 (Caltech). We thank Dr.\ T.\
Vertse for calculating the McIntyre-Haeberli phase shifts.

\narrowtext
\begin{figure}
\caption{$^3$$P_0$ $\alpha +d$ phase shift as calculated
in our different model spaces (dotted line: (i),
dot-dashed: (ii), short-dashed: (iii), and solid: (iv)).
The long-dashed line shows the phase shift as calculated
from the McIntyre-Haeberli optical potential
\protect\cite{MH}). The dots are experimental data taken
from Ref.\ \protect\cite{Jenny}.}
\label{fig1}
\end{figure}

\narrowtext
\begin{table}
\caption{Parity-nonconserving $\alpha +d$ decay widths,
calculated in the various model spaces for the $0^+$ bound
state of $^6$Li. Only the pion term is included in the
parity-violating potential.}
\begin{tabular}{ll}
Model& $\Gamma_{\alpha d}^{PNC,\pi}$  (eV) \\
\tableline
$\{\alpha_1+p+n\}$           &   $1.06\cdot 10^{-9}$ \\
$\{\alpha_1+p+n;t+h\}$       &   $4.92\cdot 10^{-10}$ \\
$\{\alpha_3+p+n\}$           &   $2.15\cdot 10^{-9}$ \\
$\{\alpha_3+p+n;t+h\}$       &   $3.55\cdot 10^{-9}$ \\
\end{tabular}
\label{tab1}
\end{table}

\narrowtext
\begin{table}
\caption{Contribution of the $(L,S)=(0,0)$ and $(1,1)$
components to the pion term of the parity-violating matrix
element. The contribution is defined by
$M_{LS}=-i\protect{\sqrt{2\pi\varrho(E_f)}} \langle
\Psi_{LS}^{^6{\rm Li}}\vert {\widehat V}^{PNC}_\pi
\vert \Psi^{\alpha d} \rangle /F_\pi$, where $\Psi^{^6{\rm
Li}}=\Psi^{^6{\rm Li}}_{00}+\Psi^{^6{\rm Li}}_{11}$ is the
total wave function in the various model spaces.}
\begin{tabular}{lr@{}lr@{}l}
Model& \multicolumn{2}{c}{$M_{00}$ $(\sqrt{eV})$} &
\multicolumn{2}{c}{$M_{11}$ $(\sqrt{eV})$}  \\
\tableline
$\{\alpha_1+p+n\}$      &\ \ \ \ --14.&5 &\ \ \ \ 44.&6 \\
$\{\alpha_1+p+n;t+h\}$  &  --12.&8 & 33.&4 \\
$\{\alpha_3+p+n\}$      &    16.&0 & 27.&0 \\
$\{\alpha_3+p+n;t+h\}$  &    32.&1 & 23.&1 \\
\end{tabular}
\label{tab2}
\end{table}

\narrowtext
\begin{table}
\caption{Parity-nonconserving $\alpha +d$ decay widths
in our full model (iv), calculated using the full
parity-violating potential (first column) and its pionic
term only (second column). The calculations have been
performed for various sets of weak coupling constants.}
\begin{tabular}{lr@{}lr@{}l}
Coupling constants&
 \multicolumn{2}{c}{$\Gamma_{\alpha d}^{PNC}$ (eV)} &
 \multicolumn{2}{c}{$\Gamma_{\alpha d}^{PNC,\pi}$ (eV)} \\
\tableline
DDH \protect\cite{DDH} & 3.&$97\cdot 10^{-9}$ &
     3.&$55\cdot 10^{-9} $ \\
DZ  \protect\cite{DZ} & 4.&$28\cdot 10^{-10}$ &
     2.&$22\cdot 10^{-10}$ \\
FCDH \protect\cite{FCDH} & 1.&$73\cdot 10^{-9}$ &
     1.&$21\cdot 10^{-9}$ \\
\end{tabular}
\label{tab3}
\end{table}


\begin{references}
\bibitem{Adelberger} E.~G. Adelberger and W.~C. Haxton,
Annu. Rev. Nucl. Part. Sci. {\bf 35} (1985) 501.
\bibitem{Holstein} W. Haeberli and B.~R. Holstein,
preprint nucl-th 9510062; to appear in {\it Symmetries in
Nuclear Physics}, edited by W.~C. Haxton and H. Henley
(1995).
\bibitem{Lobashov} V.~M. Lobashov, V.~A. Nazarenko, L.~F.
Saenko, L.~M. Smotrisky, and G.~I. Kharkevich, JETP Lett.
{\bf 5} (1967) 59; Phys. Lett. {\bf 25B} (1967) 104.
\bibitem{GFMC} B.~S. Pudliner, V.~R. Pandharipande,
J. Carlson and R.~B. Wiringa, Phys. Rev. Lett. {\bf 74}
(1995) 4396.
\bibitem{Zheng} D.~C. Zheng, J.~P. Vary and B.~R. Barrett,
Phys. Rev. {\bf C 50} (1994) 2841.
\bibitem{Tang} P.~N. Shen, Y.~C. Tang, Y. Fujiwara, and H.
Kanada, Phys. Rev. {\bf C 31} (1985) 2001; Y. Fujiwara and
Y.~C. Tang, Prog. Theor. Phys. {\bf 91} (1994) 631.
\bibitem{He6} A. Cs\'ot\'o, Phys. Rev. {\bf C 48}
(1993) 165.
\bibitem{Kok} N.~W. Schellingerhout, L.~P. Kok, S.~A.
Coon and R.~M. Adam, Phys. Rev. {\bf C 48} (1993) 2714.
\bibitem{Li6ref} B.~V. Danilin, M.~V. Zhukov, S.~N. Ershov,
F.~A. Gareev, R.~S. Kurmanov, J.~S. Vaagen, and J.~H. Bang,
Phys. Rev. {\bf C 43} (1991) 2835.
\bibitem{Kukulin} V.~I. Kukulin, V.~N. Pomerantsev,
Kh.~D. Razikov V.~T. Voronchev and G.~G. Ryzhikh,
Nucl. Phys. {\bf A586} (1995) 151.
\bibitem{Robertson} R.~G.~H. Robertson, P. Dyer, R.~C.
Melin, T.~J. Bowles, A.~B. McDonald, G.~C. Ball, W.~G.
Davies, and E.~D. Earle, Phys. Rev. {\bf C 29} (1984) 755.
\bibitem{Brown} R.~G.~H. Robertson and B.~A. Brown, Phys.
Rev. {\bf C 28} (1983) 443.
\bibitem{Burov} V.~V. Burov, V.~M. Dubovik, S.~G.
Kadmensky, Yu.~M. Tchuvil'sky, and L.~A. Tosunyan, J.
Phys. G {\bf 12} (1986) 509.
\bibitem{Varga} K. Arai, Y. Suzuki, and K. Varga, Phys.
Rev. {\bf C 51} (1995) 2488.
\bibitem{DDH} B. Desplanques, J.~F. Donoghue, and B.~R.
Holstein, Ann. Phys. (N.Y.) {\bf 124} (1980) 449.
\bibitem{Kamimura} M. Kamimura, Prog. Theor. Phys. Suppl.
{\bf 68} (1980) 236.
\bibitem{reduce} A.~T. Kruppa and A. Cs\'ot\'o,
unpublished.
\bibitem{MN} D.~R. Thompson, M. LeMere and Y.~C. Tang,
Nucl. Phys. {\bf A268} (1977) 53; I. Reichstein and
Y.~C. Tang, Nucl. Phys. {\bf A158} (1970) 529; see also
Ref.\ \protect\cite{He6}.
\bibitem{Jenny} B. Jenney, W. Gr\"uebler, V. K\"onig,
P.~A. Schmelzbach, and C. Schweizer, Nucl. Phys.
{\bf A397} (1983) 61.
\bibitem{MH} L.~C. McIntyre and W. Haeberli, Nucl. Phys.
{\bf A91} (1967) 382.
\bibitem{beta} A. Cs\'ot\'o and D. Baye, Phys. Rev. {\bf
C 49} (1994) 818.
\bibitem{DZ} V.~M. Dubovik and S.~V. Zenkin, Ann. Phys.
(NY) {\bf 172} (1986) 100.
\bibitem{FCDH} G.~B. Feldman, G.~A. Crawford, J. Dubach,
and B.~R. Holstein, Phys. Rev. {\bf C 43} (1991) 863.
\end{references}
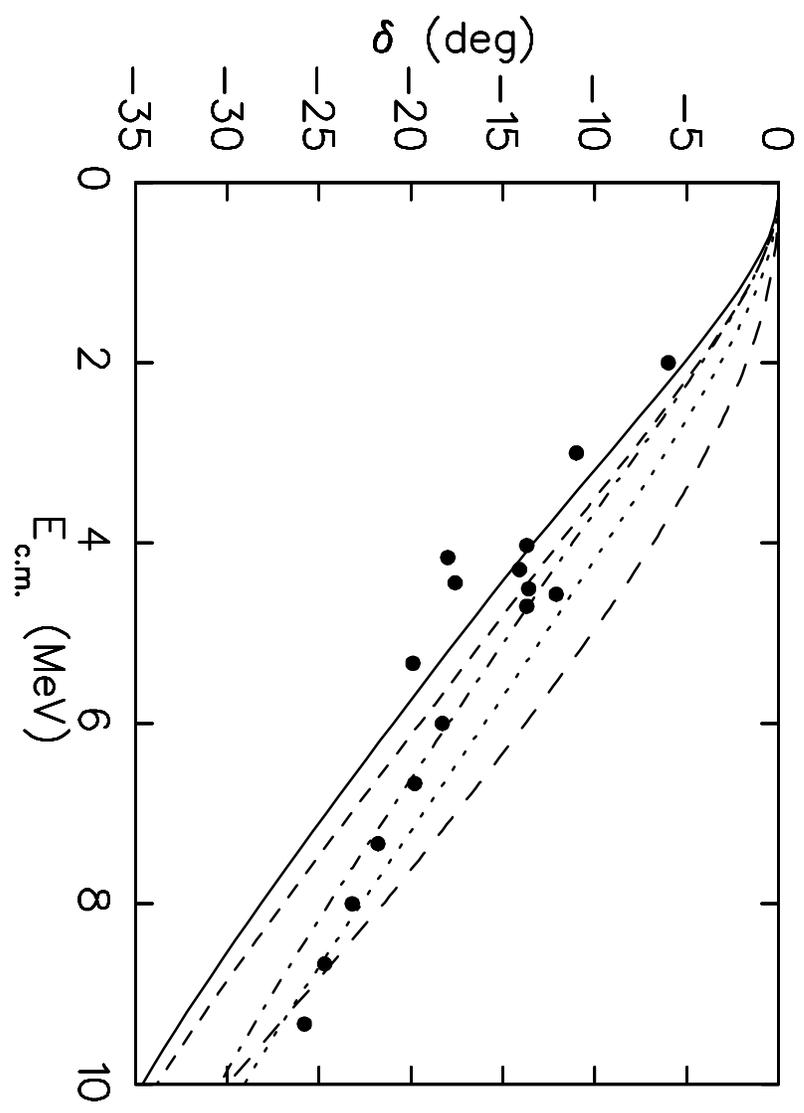
\end{document}